\documentclass[a4paper,11pt]{article}
\usepackage[utf8]{inputenc}
\usepackage[english]{babel}
\usepackage{braket}
\usepackage{xspace}
\usepackage{bbm}
\usepackage{mathdots}
\usepackage{stackrel}
\usepackage[dvipsnames]{xcolor}
\usepackage{braket,physics} 
\usepackage{dsfont}
\usepackage{appendix}
\usepackage{hyperref}
\hypersetup{
 colorlinks=true,
 linkcolor=blue,
 anchorcolor = blue,
 citecolor = blue,
 filecolor = blue,
 urlcolor = blue
}
\usepackage{mathtools}
\usepackage{bbm}
\usepackage{comment}
\usepackage{subfigure}
\usepackage[T1]{fontenc}               
\usepackage[left=3cm,
            right=2.5cm,
            top=2.5cm,
            bottom=3cm
            ]{geometry}                
\usepackage{graphicx}
\usepackage{mathrsfs}
\usepackage{appendix}
\usepackage{fancyhdr}
\usepackage{amsmath,                   
            amssymb,                   
            amsthm}                    

\usepackage{setspace}
\usepackage{cite}
\usepackage{cancel}
\usepackage[margin=30pt, bf, font=small, center, justification=justified]{caption}[2004/07/16]

\usepackage{setspace} 

\usepackage{tikz}

\title{\bf A simpler probe of the quantum Mpemba effect in closed systems}

\author{Filiberto Ares$^{1}$, Colin Rylands$^{1, 2}$,  and Pasquale Calabrese$^{1,3}$}

\date{}

\begin{document} 

\maketitle
{\small
\vspace{-5mm}  \ \\
{$^{1}$}  SISSA and INFN Sezione di Trieste, via Bonomea 265, 34136 Trieste, Italy\\[0.1cm]
\medskip
{$^{2}$} Centre for Fluid and Complex Systems, Coventry University, Coventry, CV1 2TT, United Kingdom\\[-0.1cm]
\medskip
{$^{3}$}  International Centre for Theoretical Physics (ICTP), Strada Costiera 11, 34151 Trieste, Italy\\[-0.1cm]
}

\begin{abstract}
We study the local relaxation of closed quantum systems through the relative entropy between the reduced density matrix and its long time limit. We show, using analytic arguments combined with numerical checks, that this relative entropy can be very well approximated by an entropy difference, affording a significant computational advantage. We go on to relate this to the entanglement asymmetry of the subsystem with respect to time translation invariance. In doing this, we obtain a simple probe of the relaxation dynamics of closed many-body systems and use it to re-examine the quantum Mpemba effect, wherein states can relax faster if they are initially further from equilibrium. We reproduce earlier instances of the effect related to symmetry restoration as well as uncover new cases in the absence of such symmetries. For integrable models, we obtain the criteria for this to occur using the quasiparticle picture.  Lastly, we show that, in models obeying the entanglement membrane picture, the quantum Mpemba effect cannot occur for a large class of initial states. 
\end{abstract}

\maketitle

\section{Introduction}
A closed, many-body quantum system undergoing unitary dynamics does not possess any long time limit, however, a small subsystem of it does~\cite{polkovnikov-11, gogolin-16, cem-16}. In particular, the reduced density matrix of the subsystem can relax to a stationary state.  This long time state can typically be characterized by a statistical ensemble such as a Gibbs or a generalized Gibbs ensemble~\cite{deutsch-91, srednicki-94, rdyo-07, rdo-08}. The approach to stationarity is exceedingly complex and can be probed in a myriad of ways, however certain tools are better suited to accessing universal properties of this quantum relaxation than others. One example of this is the bipartite entanglement entropy between the subsystem and its complement, which allows one to understand the relaxation through the lens of the growth of entanglement. A more recently employed tool is the entanglement asymmetry which characterizes the subsystem through its broken symmetries~\cite{amc-23}. It is defined as the  quantum relative entropy between the reduced density matrix of the subsystem, $\rho_A(t)$, and an auxiliary density matrix, $\rho_{A,G}(t)$, constructed from $\rho_A(t)$ by symmetrizing it with respect to a symmetry $G$ which acts in the subsystem~\cite{vaccaro-08, gms-09, ms-14, chmp-20, chmp-21}. Specifically, 
\begin{equation}\label{eq:symm_rho}
    \rho_{A,G}(t)=\int _G{\rm d}g \, g \rho_A(t) g^{-1},
\end{equation}
where $g$ are representations of $G$ and d$g$ an invariant  measure over the group. The entanglement asymmetry quantifies how different these two density matrices are and, as a result, how far the subsystem is from being $G$-symmetric at a given time. The asymmetry has been widely studied in recent years in many different scenarios, both in~\cite{cm-23, cv-24, fadc-24, chen-chen-24, lmac-25, frc-25, kmop-25, ampc-24, chen-25, rac-25, rac-25-2, tarabunga-24} and out of equilibrium~\cite{amvc-23, fac-24, bkccr-24, rylands-24, makc-24, carc-24, yac-24, Klobas_2024, rylands2024, yca-24,cma-24, avm-24, benini2024, bds-24, turkeshi-24, lzyz-24, krb-24, lzyzy-24-2, foligno-24, ylz-25-2, amcp-25, dgtm-25, Khor24, yu-25, gsb-25, mff-25}, revealing several intriguing phenomena. The most widely publicized of these is the quantum Mpemba effect~\cite{tblrv-25, acm-25,ylz-25}, the name given to the effect whereby a system can restore a certain symmetry faster if it initially breaks it more. This was first observed in the quench dynamics of free fermions initiated in a state which breaks particle number symmetry~\cite{amc-23}. Later, this effect was found in many other systems; in particular, its origin was understood in integrable models~\cite{rylands-24} and experimentally probed in a trapped-ion quantum simulator~\cite{joshi-24}. Other versions of the Mpemba effect have been identified in the absence of global symmetries~\cite{bhore-25} or of symmetry restoration~\cite{yca-24} and in open quantum systems, see e.g.~\cite{nf-19, cll-21, cth-23, moroder-24, ne-24, spc-25}, also experimentally~\cite{shapira, zhangexp}. 

Both the entanglement entropy and the entanglement asymmetry probe the local relaxation of a many-body quantum system in an observable independent way. The first using entanglement, the second, symmetry. A priori, there is no reason to expect any direct relationship between the two other than they deal with different aspects of the process. In this paper, we investigate this point further and show that they are related through the restoration of time translation invariance. In particular, we show that the quantum relative entropy between the reduced density matrix and its stationary value can be well approximated as the difference in the entanglement entropy of the two states. At the same level of approximation, this is the equivalent to the entanglement asymmetry where the symmetry in question is the one generated by the time evolution operator itself. We support this claim through analytic arguments and extensive numerics in a number of disparate systems.  After establishing its validity, we use this connection to investigate the quantum Mpemba effect, re-analyzing cases where local symmetries, such as particle number, are restored and then expanding the analysis beyond this to systems which do not have such symmetries. We find that previously reported instances of the quantum Mpemba effect found using the asymmetry are reproduced using the entanglement entropy, and re-derive the conditions for it to occur in integrable models. Moreover, we find new occurrences in systems which do not posses previously studied symmetries. Lastly, we show that random unitary circuits do not exhibit the Mpemba effect within the entanglement membrane picture for a large class of initial states.

\section{Relative entropy}
We want to study how far a closed quantum system is from relaxing to local equilibrium. To do this, we employ  the  quantum relative entropy
\begin{equation}\label{eq:relative_ent}
    S(\rho_A(t)||\rho_{A}(\infty))={\rm tr}[\rho_A(t)\{\log[\rho_A(t)]-\log [\rho_{A}(\infty)]\}],
\end{equation}
where $\rho_A(t)$ is the reduced density matrix of a subsystem $A$ and $\rho_A(\infty)$ is its long time limit. This quantity is always non-negative and is zero only when $\rho_A(t)=\rho_A(\infty)$, i.e. it vanishes only when the system has completely relaxed to the stationary state. The relative entropy can therefore provide direct insight into the relaxation of the system but is quite difficult to calculate both analytically and numerically. In this work, however, we show that it can be approximated to a good degree by a quantity which is much easier to treat: an  entropy difference.  

We begin by first discussing the case where the dynamics is generated by a time evolution operator which has no conserved quantities, as is the case for example in a random unitary quantum circuit. In such a scenario,  it is expected that the system relaxes locally to the infinite temperature state,
\begin{eqnarray}\label{eq:inf_temperature}
    \rho_{A}(\infty)=\frac{\mathds{1}}{2^{\ell}}~,
\end{eqnarray}
where we have assumed a lattice system with local Hilbert space of dimension $2$. In addition, the subsystem is taken to be much smaller than the full system and consist of $\ell$ contiguous sites.  It is straightforward to see that in this case the relative entropy simplifies considerably and becomes a difference of entropies,
\begin{eqnarray}\label{eq:infinite_temp}
     S(\rho_A(t)||\rho_{{A}}(\infty))=S[\rho_{{A}}(\infty)]-S[\rho_A(t)]~,
\end{eqnarray}
where $S[\rho]=-{\tr [\rho\log(\rho)]}$. The difference of entropies is typically much easier to calculate than a relative entropy but in the above scenario offers no advantage because of the simplicity of the long time state. Remarkably, however, even for non-trivial long time states and dynamics, it is possible to accurately approximate the relative entropy as an entropy difference. 

In particular, consider a system undergoing unitary Hamiltonian dynamics, so that
\begin{eqnarray}
    \rho_A(t)={\rm tr}_{\bar A}[e^{-iHt}\rho(0)e^{iHt}]
\end{eqnarray}
with $\rho(0)$ being the initial density matrix of the system and $\bar{A}$ denoting the complement of $A$. We restrict ourselves to Hamiltonians which have local interactions. Generically, it is expected that the system locally relaxes to a stationary state which is given by the diagonal ensemble reduced to the subsystem $A$~\cite{rdyo-07}. 
Namely, $\rho_A(\infty)=\rho_{{\rm d},A}$ where 
\begin{eqnarray}
    \rho_{{\rm d},A}=\sum_{\epsilon}\bra{\epsilon}\rho(0)\ket{\epsilon}{\rm tr}_{\bar{A}}\big[\ketbra{\epsilon}{\epsilon}\big] 
\end{eqnarray}
and $\ket{\epsilon}$ are the eigenstates of  $H$. We find that, in this case, the relative entropy~\eqref{eq:relative_ent} can be approximated as 
\begin{eqnarray} \label{eq:result}
S(\rho_A(t)||\rho_{{\rm{d},A}})\simeq S[\rho_{{\rm{d},A}}]-S[\rho_A(t)]~, 
\end{eqnarray}
which provides a dramatic simplification, making it amenable to the numerous analytic and numerical tools developed for calculating entanglement entropies. We note that, at the level of the full system, the relative entropy between the state and the diagonal ensemble being an entropy difference is exact, see e.g.~\cite{fgp-19}. 

\textbf{Analytic argument.} Below we provide convincing numerical evidence for the validity of our expression~\eqref{eq:result} but first we provide an analytic argument. For this we use the following property of eigenstates of local Hamiltonians~\cite{deutsch-91, akl-16, rnjk-24, ljkrn-25}, 
\begin{eqnarray}\label{eq:decomposition}
    \ket{\epsilon}\simeq \sum_{\epsilon_A}\gamma_{\epsilon_A}\ket{\epsilon-\epsilon_A}_{\bar{A}}\otimes \ket{\epsilon_A}_A
\end{eqnarray}
where $\ket{e}_{B}$ are eigenstates of the Hamiltonian $H$ restricted to act on the subsystem $B=A,\bar{A}$ with energy $e$. This approximation follows from the locality of the Hamiltonian which we can split into $H=H_A+H_{\bar{A}}+H_{A\bar{A}}$ where $H_{A\bar{A}}$ includes terms which connect $A$ and $\bar A$. In local Hamiltonians and upon taking the thermodynamic limit, we assume that the latter term can be neglected and we arrive at the above approximation.   

To obtain our simplification~\eqref{eq:result}, we start with 
\begin{eqnarray}\nonumber
\rho_A(t)&=&\sum_{\epsilon,\epsilon'}\bra{\epsilon}\rho(0)\ket{\epsilon'}\tr_{\bar{A}} \Big[ \ket{\epsilon'}\bra{\epsilon}\Big]e^{-i (\epsilon'-\epsilon)t}\\\nonumber
    &\simeq& \sum_{\epsilon,\epsilon',\varepsilon,\varepsilon'}\bra{\epsilon}\rho(0)\ket{\epsilon'}\gamma_{\varepsilon'}\gamma^{\star}_{\varepsilon}\tr_{\bar{A}} \Big[ \ket{\epsilon'-\varepsilon'}_{\bar{A}}\bra{\epsilon-\varepsilon}_{\bar{A}}\otimes  \ket{\varepsilon'}_{A}{}\bra{\varepsilon} _{A}\Big]e^{-i (\epsilon'-\epsilon)t}\\\label{eq:rhoA_approx}
&=&\sum_{\epsilon,\varepsilon,\varepsilon'}\bra{\epsilon}\rho(0)\ket{\epsilon-\varepsilon+\varepsilon'}\gamma_{\varepsilon'}\gamma^{\star}_{\varepsilon}\ket{\varepsilon'}_{A}\bra{\varepsilon}_{A} e^{-i (\varepsilon'-\varepsilon)t}~,
\end{eqnarray}
where in the first line we have expanded $\rho(t)$ in the basis of $H$ and in going to the second we have used 
\eqref{eq:decomposition}. We then note that using the decomposition of energy eigenstates written above,~\eqref{eq:decomposition}, we have that 
\begin{eqnarray}
    \rho_{{\rm d}, A}\simeq\sum_{\varepsilon}\Pi^A_{\varepsilon}  \rho_{ A}\Pi^A_{\varepsilon},
\end{eqnarray}
where $\Pi^A_{\varepsilon}=\ket{\varepsilon}_{A}\bra{\varepsilon}_{A}$ is a projector onto the eigenstate of $H_A$ with energy $\varepsilon$. This expression states that the reduced diagonal ensemble is a stationary state of the dynamics generated by the subsystem Hamiltonian.  Armed with the expression for $\rho_A(t)$ in Eq.~\eqref{eq:rhoA_approx} and the stationarity of $\rho_{{\rm d},A}$ we return to the relative entropy~\eqref{eq:relative_ent}. 
Focusing on the second term, we find
\begin{eqnarray}
    {\rm tr}[\rho_A(t)\log\rho_{{\rm d},A}]&\simeq&\sum_{\varepsilon}  {\rm tr}[\rho_A(t) \Pi_{\varepsilon}^A\log\{\rho_{{\rm d},A}\}\Pi_{\varepsilon}^A]\\
    &=&\sum_{\varepsilon}  {\rm tr}[\Pi_{\varepsilon}^A\rho_A(t) \Pi_{\varepsilon}^A\log\{\rho_{{\rm d},A}\}]\\
    &\simeq& {\rm tr}[\rho_{{\rm d},A} \log\{\rho_{{\rm d},A}\}],
\end{eqnarray}
where in the last line we have used the expression~\eqref{eq:rhoA_approx}. Inserting this result in~\eqref{eq:relative_ent}, we arrive at~\eqref{eq:result}, justifying our statement. 

\textbf{Connection with asymmetry.} We note that, within this approximation, the relative entropy is equivalent to the entanglement asymmetry, denoted $
\Delta S_A(t)$,  using the Hamiltonian as the charge, i.e. the entanglement asymmetry with respect to time translation invariance. Explicitly, 
\begin{eqnarray}
    S(\rho_A(t)||\rho_{{\rm{d},A}})\simeq \Delta S_A(t)= {\rm tr}[\rho_A(t)\{\log[\rho_A(t)]-\log[\rho_{A,H}(t)]\}],
\end{eqnarray}
where
\begin{eqnarray}
    \rho_{A,H}(t)=\sum_{\varepsilon}\Pi^A_{\varepsilon}\rho_A(t)\Pi^A_{\varepsilon}.
\end{eqnarray}
The equivalence to Eq.~\eqref{eq:symm_rho} arises from the fact that time translations form a non-compact Lie group, whose elements are $
e^{-i\tau H_A 
}$, $
\tau \in \mathbb{R}$.  Thus the relative entropy measures how close the local system is to stationarity which, quite naturally, via the entanglement asymmetry is equivalent to how close the state is to being locally time translationally invariant. 

\begin{figure}
\includegraphics[width=0.5\textwidth]{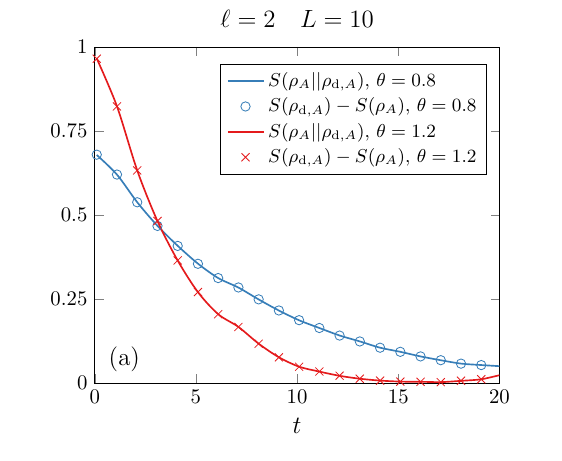}
\includegraphics[width=0.5\textwidth]{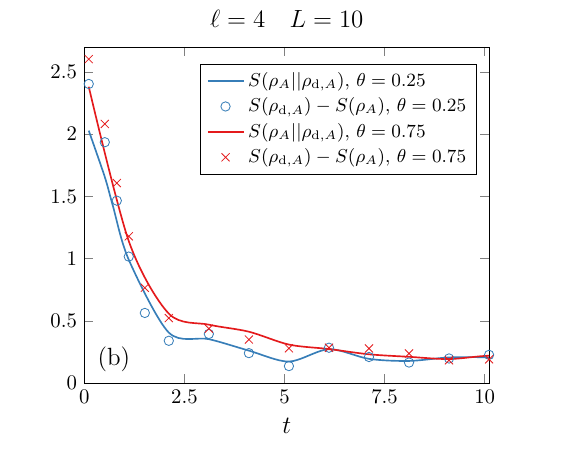}
\includegraphics[width=0.5\textwidth]{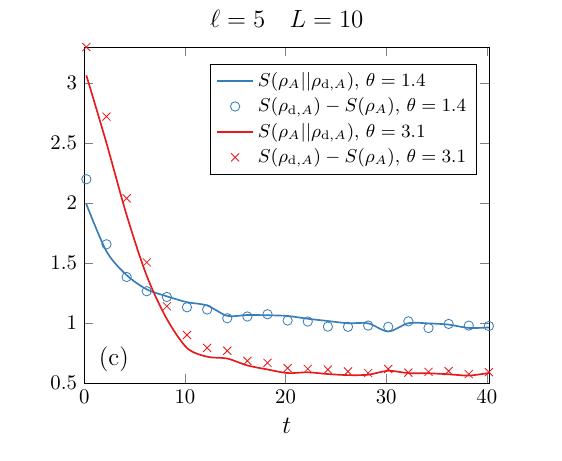}
\includegraphics[width=0.5\textwidth]{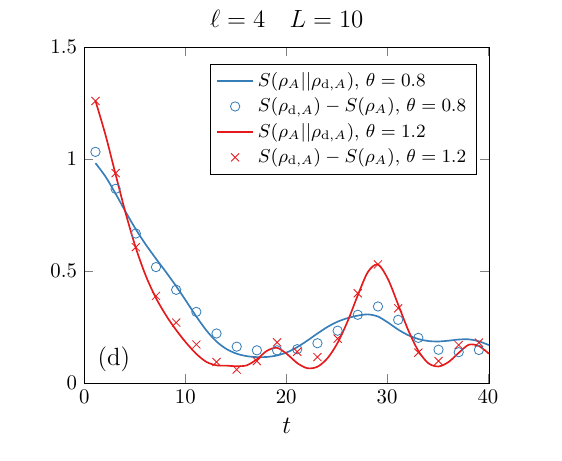}
\caption{Numerical check of the approximation in Eq.~\eqref{eq:result}. We compare the relative entropy (solid lines) and the difference between entanglement entropies (symbols) of $\rho_A(t)$ and the diagonal ensemble $\rho_{{\rm d}, A}$ in a subsystem $A$ of $\ell$ contiguous sites.
Both quantities have been obtained with exact diagonalization.
(a) Quench in the XXZ spin-$1/2$~\eqref{eq:xxz} with $\Delta=0.5$ and $h=0.1$ from two tilted ferromagnetic configurations~\eqref{eq:ferro} (b) Quench in the spin-$1/2$ chain described by the next-nearest neighbor Hamiltonian~\eqref{eq:nnn} with $\Delta=1.5$, $\Delta_2=0.5$, and $J_2=1$ from two tilted N\'eel states~\eqref{eq:neel}. (c) Quench in the Mixed-Field Ising chain~\eqref{eq:mf_ising} with the parameters indicated in the main text and prepared in different tilted ferromagnetic states~\eqref{eq:ferro}. (d) Quench in the long-range XX spin-$1/2$ chain~\eqref{eq:lr_xx} with $\alpha=1$ and $J_0=1/4$, starting from two tilted ferromagnetic states~\eqref{eq:ferro}.} 
\label{fig:check}
\end{figure}

\textbf{Numerical checks.} In Fig.~\ref{fig:check}, we numerically check the approximation in Eq.~\eqref{eq:result}. We compute both sides of that equation with exact diagonalization in different quenches of a spin-1/2 chain of length $L$ and taking open boundary conditions. Solid lines correspond to the relative entropy $S(\rho_A(t)||\rho_{{\rm d}, A})$ and the symbols to the difference between the entropies  $S(\rho_{{\rm d}, A})-S(\rho_A(t))$ for a subsystem $A$ of $\ell$ contiguous spins. We obtain in all cases a very good agreement, even for small subsystem sizes.

In panel (a), the chain is initialized in two different tilted ferromagnetic states,
\begin{equation}\label{eq:ferro}
    \ket{{\rm F}, \theta}=e^{-i\theta/2\sum_{j=1}^L\sigma_j^y}\ket{\uparrow \cdots \uparrow},
\end{equation}
and evolves unitarily with the XXZ Hamiltonian
\begin{equation}\label{eq:xxz}
 H_{\rm XXZ}=-\frac{1}{4}\sum_{j=1}^{L-1}(\sigma_j^x\sigma_{j+1}^x+\sigma_j^y\sigma_{j+1}^y+\Delta \sigma_j^z\sigma_{j+1}^z)+\frac{h}{2}\sum_{j=1}^L\sigma_j^z.
 \end{equation}
 This is one of the most well studied interacting spin chain models. Here $
 \Delta$ is the anisotropy which breaks the $SU(2)$ invariance of the model and $h$ is a global magnetic field applied in the $z$ direction. From the figure, we see that the agreement between the relative entropy and the difference of entropies is excellent even at such small system and subsystem sizes. This Hamiltonian is a paradigmatic example of an interacting integrable model and so to investigate whether integrability plays any role in the validity of~\eqref{eq:result} we study other systems which break integrability. 
 
 In panel (b), the chain is initially in a tilted N\'eel state, 
 \begin{equation}\label{eq:neel}
    \ket{{\rm N}, \theta}=e^{-i\theta/2\sum_{j=1}^L\sigma_j^y}\ket{\uparrow \downarrow \cdots \uparrow \downarrow},
\end{equation}
and the post-quench Hamiltonian is
\begin{equation}\label{eq:nnn}
H_{\rm NNN}=H_{\rm XXZ}-\frac{J_2}{4}\sum_{j=1}^{L-2}\left(\sigma_j^x\sigma_{j+2}^x+\sigma_{j}^y\sigma_{j+2}^y+\Delta_2\sigma_{j}^z\sigma_{j+2}^z\right),
\end{equation}
i.e. we add next-nearest neighbor interactions to the XXZ spin chain Hamiltonian~\eqref{eq:xxz}, which breaks the integrability of the model. Once again the agreement is excellent. 

Both Hamiltonians~\eqref{eq:xxz} and \eqref{eq:nnn} commute with the transverse magnetization, $Q=\sum_j\sigma_j^z$. The $U(1)$ symmetry generated by $Q$ is instead broken in the initial states~\eqref{eq:ferro} and~\eqref{eq:neel}. The occurrence of the Mpemba effect in the restoration of this symmetry in these models has been studied with the entanglement asymmetry and applying exact diagonalization in Ref.~\cite{amc-23}. 

In panel (c), we consider quenches from tilted ferromagnetic states~\eqref{eq:ferro} in a system that lacks any symmetry except energy conservation, the mixed-field Ising chain, 
\begin{equation}\label{eq:mf_ising}
H_{\rm Ising}=\sum_{j=1}^{L-1}\sigma_j^z\sigma_{j+1}^z+g\sum_{j=1}^{L}\sigma_j^x+h\sum_{j=2}^{L-1}\sigma_j^z + h' \sigma_1^z-h'\sigma_L^z.
\end{equation}
We take different and opposite magnetic field $h'$ in the $z$-axis at the boundaries to break the reflection symmetry. We choose $g=(\sqrt{5}+5)/8$ and $h=(\sqrt{5}+1)/4$, for which the model is chaotic~\cite{bch-11, kh-13, zkh-15}, and $h'=1/4$. The occurrence of the quantum Mpemba effect in this quench for the same parameters has been recently examined in Ref.~\cite{bhore-25} using the trace distance between $\rho_{A}(t)$ and $\rho_{{\rm d}, A}$. 

Finally, in panel (d), we study quenches in a long-range XX spin chain, 
\begin{equation}\label{eq:lr_xx}
H_{\rm LR}=\sum_{j<j'}\frac{J_0}{4|j-j'|^\alpha}\left(\sigma_{j}^x\sigma_{j'}^x+\sigma_{j}^y\sigma_{j'}^y\right),
\end{equation}
starting from different tilted ferromagnetic states~\eqref{eq:ferro}. This is the setup used in Ref.~\cite{joshi-24} to experimetally observe in a trapped-ion quantum simulator the quantum Mpemba effect. There the effect was probed by directly measuring the entanglement asymmetry associated with $Q$ and the Frobenius distance between 
$\rho_A(t)$ and $\rho_{{\rm d}, A}$ with the randomized measurement toolbox~\cite{elben-23}.  
In this case, the analytic reasoning presented above to justify the approximation in Eq.~\eqref{eq:result} is not valid since it entirely relies on the locality of the interactions of the Hamiltonian. Nevertheless, as we can conclude from the plot, Eq.~\eqref{eq:result} is also here a good approximation. 

\textbf{Quantum Mpemba Effect.}
Using our new expression~\eqref{eq:result}, we can explore the quantum Mpemba effect. In terms of the entanglement asymmetry, this effect was defined as occurring if an initial state which breaks a symmetry more, can restore it faster than one which initially breaks it less. One can view the symmetry breaking and eventual restoration as a proxy for being far from or close to local relaxation. Thus, more generally the quantum Mpemba effect is the statement that an initial state which is further from equilibrium can relax faster than one which is closer to it. 

More specifically, consider two different initial states, described by density matrices $\rho^{1,2}$, the first state is said to be further from local equilibrium if 
\begin{eqnarray}
S(\rho^1_A(0)||\rho^1_{{\rm d},A})>S(\rho^2_A(0)||\rho^2_{{\rm d},A})
\end{eqnarray}
and the quantum Mpemba effect occurs if there exists a time $t_M$ such that 
\begin{eqnarray}
    S(\rho^1_A(t)||\rho^1_{{\rm d},A})<S(\rho^2_A(t)||\rho^2_{{\rm d},A})
~~\forall t>t_M.
\label{condlt}
\end{eqnarray}
For lowly entangled  initial states, the first condition for the quantum Mpemba effect becomes 
\begin{eqnarray}\label{eq:cond2int}
S[\rho^1_{{\rm d},A}]>S[\rho^2_{{{\rm d},A}}]. 
\end{eqnarray}
Recalling that $S[\rho^{1,2}_{{\rm d},A}]$ is also the thermodynamic entropy of $\rho^{1,2}_{{\rm d},A}$, we conclude that the content of this condition is that the  state $\rho^1$ is more delocalized among the eigenstates of $H_A$ than $\rho^2$. This is similar in spirit to the approach of Ref. \cite{bhore-25} where the inverse participation ratio of the initial state with respect to the full Hamiltonian $H$ is used to understand the quantum Mpemba effect. 

In Fig.~\ref{fig:check}, we can observe some instances where the quantum Mpemba effect occurs. In panel (a), at $t=0$, $S(\rho_A(t)||\rho_{{\rm d}, A})$ is larger for $\theta=1.2$ than for $\theta=0.8$. After the quench, the relative entropy decreases faster for $\theta=1.2$, eventually crossing that of $\theta=0.8$, and $\rho_A(t)$ approaches $\rho_{{\rm d}, A}$ sooner. The same happens in panels (c) and (d). The crossing is the main signature of the quantum Mpemba effect. Instead, in panel (b), the relative entropies for any two pairs of initial configurations~\eqref{eq:neel}  with different angle $\theta$ do not cross and there is not Mpemba effect.
As we have just mentioned, the quantum Mpemba effect has been already studied in the quenches considered in the figure using the entanglement asymmetry or a distance between $\rho_A(t)$ and $\rho_{{\rm d}, A}$. Compared to those observables, $S(\rho_A(t)||\rho_{{\rm d}, A})$ displays a similar qualitative behavior and detects the quantum Mpemba effect in the same cases considered here.

\section{Quantum Mpemba effect in integrable models}\label{sec:Integrable_models}

Our expression~\eqref{eq:result} for the relative entropy is much easier to calculate than the un-approximated form or indeed the entanglement asymmetry. Using it we can analytically access many different quench protocols. In particular, for integrable models, the quasiparticle picture~\cite{cc-05, ac-17, ac-18,c-20} allows one to obtain analytic expressions for the entanglement entropy. From this, we find that in the thermodynamic limit and for large subsystem size $\ell$
\begin{eqnarray}\label{eq:qp}
S(\rho_A(t)||\rho_{{\rm d},A})\simeq\sum_m\int{\rm d}k \rho^t_m(k)[\ell -\min(2 |v_m(k)|t,\ell)]s\{n_m(k)\},
\end{eqnarray}
where the sum is over different quasiparticle species, which are indexed by $m$,  $v_m(k)$ are the velocities of the quasiparticles in the system, $s[x]=-x\log x-(1-x)\log(1-x)$, and  finally, $\rho^t_m(k)$ and $n_m(k)$ are the total density of states and occupation function for the species $m$. 

We can expand Eq.~\eqref{eq:qp} about its initial value and see that it decays linearly with $t$. The slope at short times being $ 2
\sum_m \int {
\rm d
}k \rho^t_m|v_m|s\{n_m\}$, however the existence of the Mpemba effect cannot be inferred from the short time behaviour \cite{carc-24}. Therefore, we should expand about the stationary value at long time and understand how the system approaches local equilibrium. To do this, we simply expand the integral about the slowest quasiparticle modes. If we assume that the velocity $v_m(k)$ has a single zero at $k=0$, then we can expand it as $v_m(k)\approx v_m^{(1)} k$. Furthermore we expand the mode occupation as $n_m(k)\approx n_m^{(0)}+n^{(2)}_m k^2$,  and retain only the lowest order of the density of states $\rho^t_m(k)\approx\rho^{t\, (0)}_m$ which is exact in the free case since then density of states is $k$ independent. The late time approach to equilibrium is then given by 
\begin{eqnarray}\label{eq:long_time}
 S(\rho_A(t)||\rho_{{\rm d},A}) \simeq\begin{cases}
     \sum_m \frac{\rho_m^{t (0)}s[n_m^{(0)}]}{2 v_m^{(1)}}\frac{\ell^2}{t}~~& ~~ n_m^{(0)}\neq 0\\ \\
     
     \sum_m \frac{\rho_m^{t (0)} n_m^{(2)}}{ 6[v_m^{(1)}]^3}\frac{\ell^4}{t^3} \log \big(\frac{t}{ \ell}\big)~~& ~~ n_m^{(0)}= 0 ~,~\forall m
 \end{cases}
\end{eqnarray}
The second scenario is more common, but the first can be realized when quenching, for example, from the Neel state or from a ground state of a critical model. 
It is worth noting that this large-time behavior is fully consistent with the results obtained for the entanglement asymmetry in Ref.~\cite{rylands-24}.


We now turn our attention to the occurrence of the quantum Mpemba effect.
The condition involving the initial value of the relative entropy, given in Eq.~\eqref{eq:cond2int}, has already been discussed above, as it does not depend on the integrability of the system.
The second condition~\eqref{condlt}, i.e. the large time behavior, can be rewritten in terms of initially occupied modes and quasiparticle velocities by inspecting the long time limit given above in Eq.~\eqref{eq:long_time}. In the simplest case, where there is only a single quasiparticle species and both states have $n^{(0)}=0$, i.e. no occupation of the $k=0$ mode, this second condition \eqref{condlt} becomes
\begin{eqnarray}\label{eq:cond}
    \frac{n^{(2)} \rho^{t (0)}}{[v^{(1)}]^3}\Bigg|_{\rho^1}<\frac{n^{(2)}\rho^{t (0)}}{[v^{(1)}]^3}\Bigg|_{\rho^2}~.
\end{eqnarray}
Note that in free models the quasiparticle velocity is fixed by the Hamiltonian and $\rho^t=1/2\pi$, so the condition reduces to a statement concerning the occupation of the slowest modes. In interacting integrable models, however, the velocity is dressed by the interactions in a state dependent way 
\cite{cdy-2016,bcdf-16} and can be different on both sides.  On the other hand, if both states have $n^{(0)}$ non zero, then the second condition~\eqref{condlt} can be rewritten as
\begin{eqnarray}
   \frac{s[n^{(0)}] \rho^{t (0)}}{v^{(1)}}\Bigg|_{\rho^1}<\frac{s[n^{(0)}]\rho^{t (0)}}{v^{(1)}}\Bigg|_{\rho^2}.
\end{eqnarray}
As before, in free cases this inequality becomes a statement about the entanglement of the slowest quasiparticles but is more complicated in the presence of interactions.  Finally, if state $1$ has $n^{(0)}= 0$ while state 2 has $n^{(0)}\neq 0$, then the quantum Mpemba effect will always occur, while the converse is true: If state 1 has $n^{(0)}\neq 0$  but state 2 does, then there can be no quantum Mpemba effect. We now look at some specific examples.  

\textbf{XY spin chain.} Let us first consider quenches in the XY spin-$1/2$ chain,
\begin{equation}\label{eq:xy}
H_{\rm XY}=-\sum_{j}\left(\frac{1+\gamma}{4}\sigma_{j}^x\sigma_{j+1}^x+\frac{1-\gamma}{4}\sigma_j^y\sigma_{j+1}^y\right)+\frac{h}{2}\sum_{j}\sigma_j^z.
\end{equation}
This Hamiltonian can be mapped through the Jordan-Wigner transformation to a quadratic fermionic chain which can be exactly solved by performing a Bogoliubov rotation. The evolution of the entanglement entropy in a sudden quench from the ground state of~\eqref{eq:xy} with couplings $(h_0, \gamma_0)$ to another set $(h, \gamma)$ was calculated \textit{ab initio} in Ref.~\cite{fc-08}. In this case, the quasiparticle velocity is given by $v(k)=\partial_k\epsilon (k)$, where $\epsilon(k)$ is the dispersion relation of the post-quench Hamiltonian, 
$\epsilon(k)=\sqrt{(h-\cos(k))^2+\gamma^2\sin^2(k)}$.  The mode occupation in the stationary state is
\begin{equation}
n(k)=\frac{1-\cos\Delta_k}{2}
\end{equation}
and
\begin{equation}
\cos\Delta_k=\frac{hh_0 - (h+h_0)\cos(k)+ \cos^2(k) + \gamma\gamma_0\sin^2(k)}{\epsilon(k)\epsilon_0(k)},
\end{equation}
where $\epsilon_0(k)$ is the dispersion relation of the pre-quench Hamiltonian, with parameters $(h_0, \gamma_0)$.

\begin{figure}[t]
\includegraphics[width=0.5\textwidth]{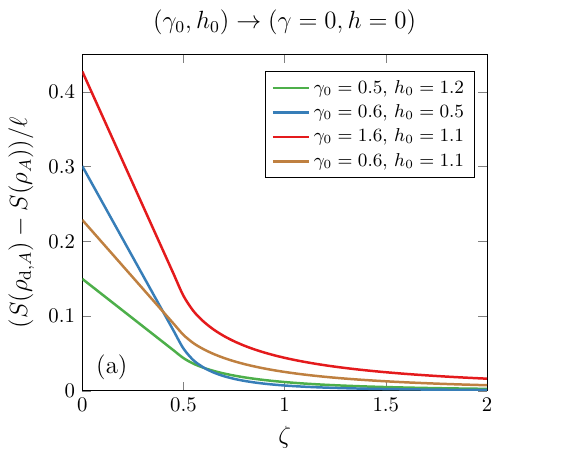}
\includegraphics[width=0.5\textwidth]{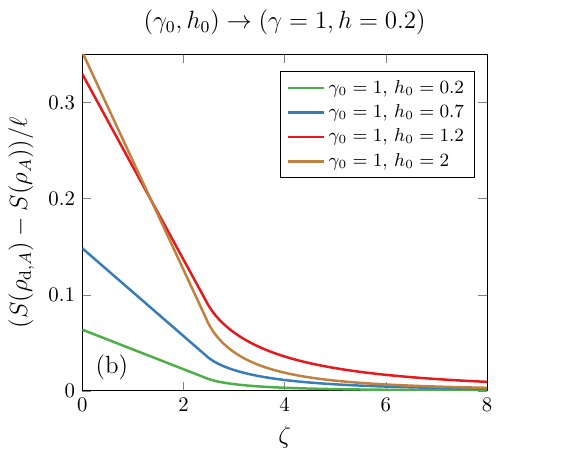}
\includegraphics[width=0.5\textwidth]{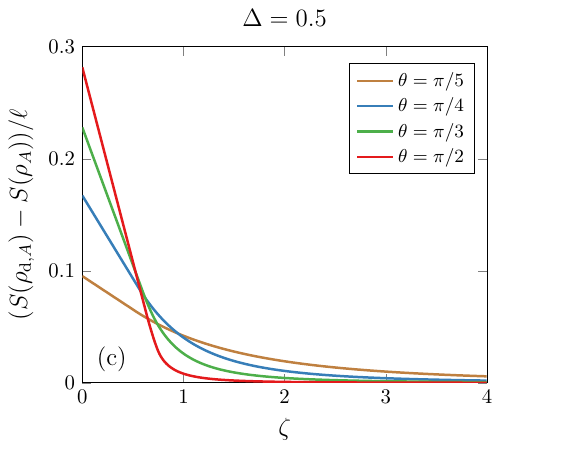}\hspace{-0.5cm}
\includegraphics[width=0.5\textwidth]{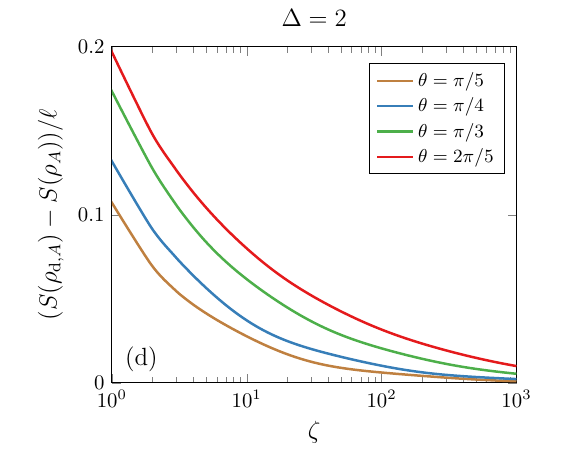}
\caption{Time evolution of the relative entropy between $\rho_A(t)$ and $\rho_{{\rm d}, A}$ after different quenches in integrable models as a function of $\zeta=t/\ell$. The solid curves in all panels were obtained using the approximation of Eq.~\eqref{eq:result} and the quasiparticle picture~\eqref{eq:qp}. (a) Quench from different ground states of the XY spin chain~\eqref{eq:xy} with parameters $(\gamma_0, h_0)$ to the XX spin chain $(\gamma=0, h=0)$. (b) Quench in the quantum Ising chain $(\gamma_0=\gamma=1)$ from several values of the external magnetic field $h_0$ to $h=0.2$. (c-d) Quench starting from tilted ferromagnetic states~\eqref{eq:ferro} with tilting angle $\theta$ in the XXZ spin chain~\eqref{eq:xxz}, taking $h=0$ and $\Delta=0.5$ (gapless chain, panel (c)) and $\Delta=2$ (gapped chain, panel (d)). } 
\label{fig:qp}
\end{figure}

In Fig.~\ref{fig:qp}~(a), we analyze the quench from the ground state of~\eqref{eq:xy} for different $(h_0, \gamma_0)$ to the Hamiltonian with $h=0$ and $\gamma=0$, which corresponds to the XX spin chain. This Hamiltonian commutes with the transverse magnetization,  $Q=\sum_j\sigma_j^z$, or particle number in fermionic language, which is otherwise broken for $\gamma\neq 0$. The quantum Mpemba effect in this quench was studied in Ref.~\cite{makc-24} through the restoration of the $U(1)$ symmetry generated by $Q$ using the entanglement asymmetry. In Fig.~\ref{fig:qp}~(a), we take the same initial ground states as in Fig. 7 (top) of Ref.~\cite{makc-24}. The quantum Mpemba effect occurs for the same pairs of initial states as when entanglement asymmetry is used.  

In Fig.~\ref{fig:qp}~(b), we consider quenches on the line $\gamma=1$, which corresponds to the quantum Ising chain, changing $h_0$ to $h<1$. Here both the pre- and post-quench Hamiltonians break the $U(1)$ particle number symmetry and we cannot use the corresponding entanglement asymmetry to study the quantum Mpemba effect. As one can see in the plot, and check from the conditions in Eq.~\eqref{eq:cond}, the quantum Mpemba effect occurs for pairs of initial ground states in the paramagnetic phase $(h>1)$ quenched to the ferromagnetic phase ($0\leq h<1$). 

\textbf{XXZ Spin Chain.} We now examine an interacting integrable model and use the previously introduced XXZ chain \eqref{eq:xxz} quenched from the tilted ferromagnetic state 
\eqref{eq:ferro}. The entanglement dynamics in this scenario were studied in~Ref.~\cite{ac-17} and the dynamics of the asymmetry investigated in Ref.~\cite{rylands2024}. In the latter, it was found that the quantum Mpemba effect, with respect to the restoration of the initially broken magnetization symmetry,  occurs for $\Delta <1$. In Fig.~\ref{fig:qp}~(c), we plot the entropy difference for several different tilt angles with $
\Delta=0.5$ and see that all lines cross. Thus the  Mpemba effect is seen also using the new probe. For $\Delta>1$, it was observed that no Mpemba effect occurs and moreover the relaxation to the stationary state is significantly delayed. This is seen also in Fig.~\ref{fig:qp}~(d) which exhibits no crossing of the lines and a much slower rate of decay of the difference. The underlying reason for this is the same as for the entanglement asymmetry. The $
\Delta>1$ regime contains an infinite number of quasiparticles species which can be thought of as bound states of all sizes. The larger the bound state, the slower it is, leading to a much slower relaxation. In the fully tilted case, $\theta=\pi/2$ the transport in the system is diffusive as opposed to ballistic for $\theta<\pi/2$. Thus, the fully tilted case will always relax slower than any other tilt angle, preventing the Mpemba effect. In contrast, for $
\Delta<1$ the model contains only a finite number of bound states and the transport is always ballistic.

\section{Quantum Mpemba effect in random unitary circuits}

In Section \ref{sec:Integrable_models}, we discussed models whose entanglement dynamics obeys the quasiparticle picture. As a counterpoint to this, we now briefly discuss models which do not possess quasiparticle excitations and moreover have no symmetries at all. 
As mentioned above, these systems locally relax to an infinite temperature state~\eqref{eq:inf_temperature} and the relative entropy is exactly the entropy difference~\eqref{eq:infinite_temp}. We focus specifically on brickwork random unitary circuits~\cite{fknv-23} and employ two standard techniques when studying the entanglement dynamics of such models. 

\begin{figure}[t]
\centering
\includegraphics[width=0.49\textwidth]{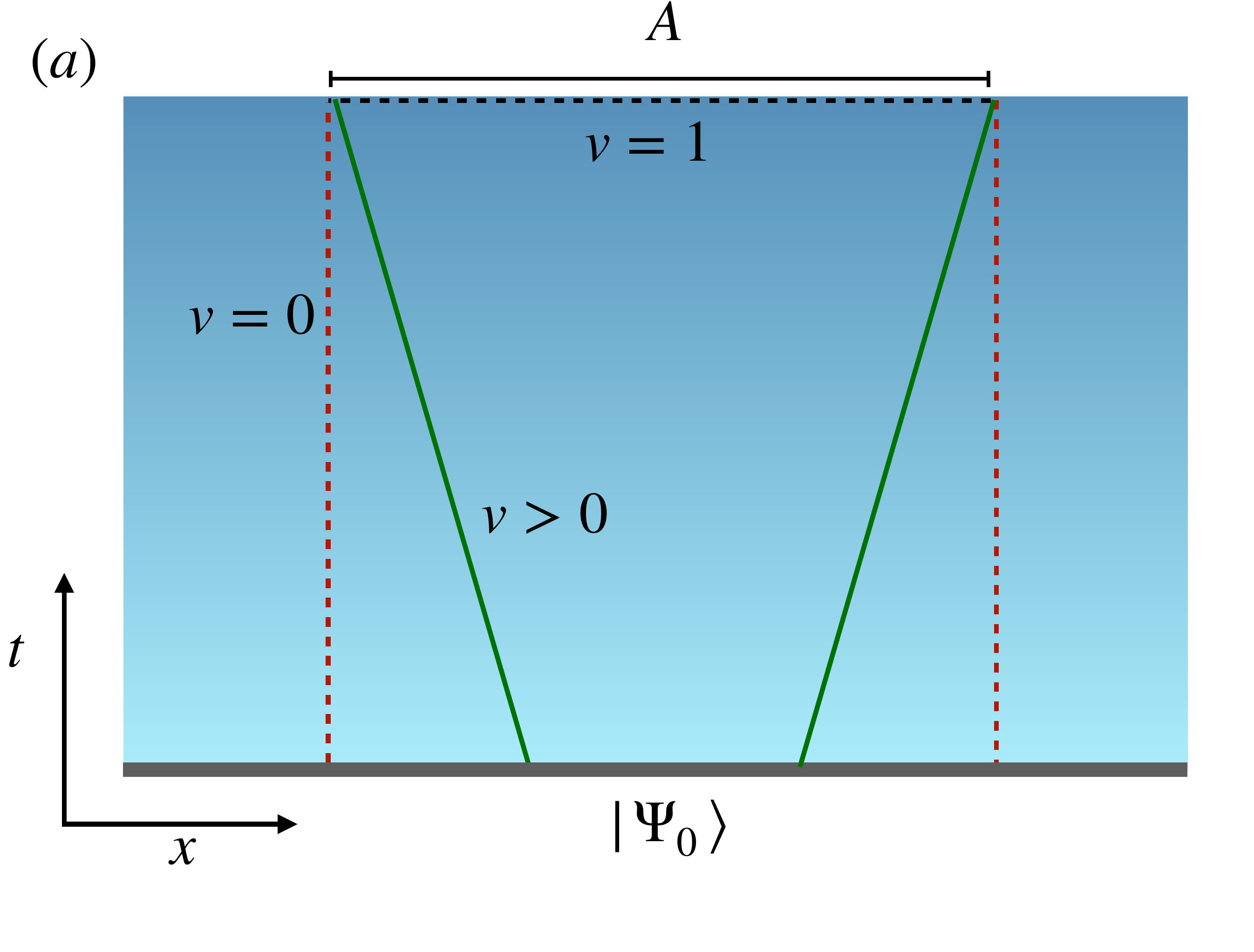}
\includegraphics[width=0.49\textwidth]{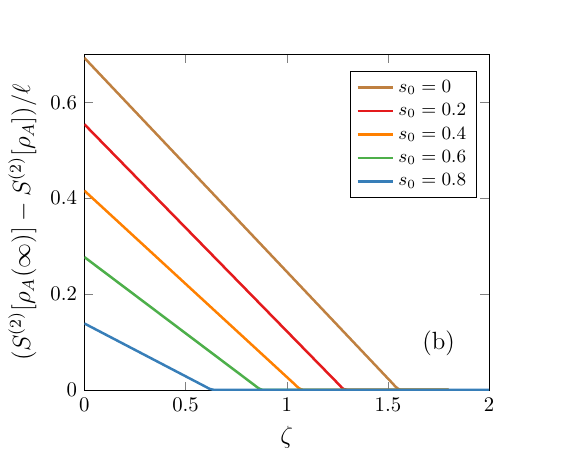}
\caption{(a) Schematic representation of the membrane picture. (b) Difference between the R\'enyi-2 entropies of $\rho_A(t)$ and $\rho_A(\infty)$ as a function of $
\zeta=t/
\ell$ in a brickwork random unitary circuit initialized in a state with R\'enyi-2 entanglement entropy $s_0\log(q)\ell$ and $q=2$. The curves are the prediction of the membrane picture~\eqref{eq:membrane}, taking as the membrane tension Eq.~\eqref{eq:tension}.} 
\label{fig:ruc}
\end{figure}

The first is to generalize the difference of von Neumann entropies to the difference of R\'enyi entropies via
\begin{eqnarray}
    S[\rho_A(t)]\to S^{(n)}[\rho_A(t)]=\frac{1}{1-n}\log[\tr_{A}\rho_A^n(t)]
\end{eqnarray}
and similarly for $S[\rho_{{\rm d},A}]\to S^{(n)}[\rho_{{\rm d},A}] $. The difference between these two quantities is no longer a relative entropy, nevertheless we expect that $S^{(n)}[\rho_{{\rm d},A}] -S^{(n)}[\rho_{A}(t)] $ is a good indicator of the behaviour of $S(\rho_A(t)||\rho_{{\rm d},A})$. The second technique we use is the entanglement membrane picture~\cite{jhn-18, zn-20}. This coarse grained effective theory posits that entanglement is not generated by the propagation of quasiparticles produced by the initial state, but rather, locally along spacetime  trajectories. These paths emanate from the edges of $A$ at time $t$ and  either terminate on the initial state at $t=0$ or join together linking both edges in a single path. The entropy is found by summing over all such paths, however in practice a single configuration will dominate in a given time window.  
The membrane picture result for the entropy is  
\begin{align}\label{eq:membrane}
    S^{(n)}[\rho_A(t)]={\rm min}[ s_0^{(n)}(\ell, vt)+2t \mathcal{E}^{(n)}(v), \ell \log(q)]~,
\end{align}
where $q$ is the local Hilbert space dimension of the circuit. One sees that, similar to the quasiparticle picture~\eqref{eq:qp}, it is the minimum of two distinct terms which now correspond to different path configurations. The first term corresponds to the situation where there are two paths starting from the edges of $A$ and ending on the initial state. It receives two contributions, $\mathcal{E}^{(n)}(v)$, which describes the entropy production along the two paths (we assume the configuration to be reflection symmetric), and $s^{(n)}_0(x)$, which is the contribution to the entanglement from the initial state. This depends on $0\leq v\leq 1$ which is the slope of the paths, $v=0$ being a purely time-like or vertical  path while $v=1$ being space-like or horizontal (see Fig.~\ref{fig:ruc}~(a)). The second term corresponds to the configuration where the path proceeds across the subsystem and connects the two edges without touching the initial state, i.e. $v=1$.  

For a lowly entangled state, $s^{(n)}_0=0$ and the slope in the early time regime is fixed to be $v=0$, independently of any other properties of the initial state. Thus within the membrane picture, random unitary dynamics of entanglement from lowly entangled states are indistinguishable from one another, thereby ruling out any QME. For highly entangled initial states the slope $v$ depends on the initial state, thus leaving room for a QME to occur. As an example, we take $n=2$ and study volume law entangled initial states, 
\begin{eqnarray}
    s^{(2)}_0 (\ell,vt)=s_0 \log(q) |\ell-2vt|~.
\end{eqnarray}
with $0\leq s_0\leq 1$. In this case the analytic form of $\mathcal{E}^{(2)}$ is known \cite{jhn-18},
\begin{eqnarray}\label{eq:tension}
    \mathcal{E}^{(2)}(v)=\log\Big[\frac{q^2+1}{q}\Big]+\Big[\frac{1+v}{2}\Big]\log \Big[\frac{1+v}{2}\Big]+\Big[\frac{1-v}{2}\Big]\log\Big[\frac{1-v}{2}\Big],
\end{eqnarray}
where the slope is given by 
\begin{eqnarray}
v=\frac{q^{s_0}-1}{q^{s_0}+1}~.
\end{eqnarray}
In Fig.~\ref{fig:ruc}, we plot $S^{(2)}[\rho_A(\infty)]-S^{(2)}[\rho_A(t)]$ for $q=2$ and different values of $s_0 $. We see that the larger the value of $s_0$ the closer the state is to equilibrium while at the same time the  relaxation rate is slower. Despite this, however there is also no quantum Mpemba effect as the states with larger $s_0$ always relax quicker.  

We note that, when the system relaxes to the infinite temperature state, not only is the relative entropy exactly equal to the entropy difference but differences of R\'enyi entropies can be exactly related to a series of distances between density matrices.
 In particular, one can show that 
 \begin{eqnarray}
 \frac{ ||\rho_A(t)-\rho_A(\infty)||^2_2}{||\rho_A(\infty)||^2_2}=
 \left[e^{S^{(2)}[\rho_A(\infty)]-S^{(2)}[\rho_A(t)]}-1\right]
 \end{eqnarray}
 where $||O||^2_2=\tr[OO^\dagger]$  is the square of the Frobenius norm.  Unlike the relative entropy, the Frobenius distance is a genuine distance in the space of density matrices and so we expect its behaviour to be indicative of all other distances. By going to higher order R\'enyi entropies, one could also study all Schatten $n-$norms, however  no explicit formula exists  for $
\mathcal{E}^{(n)}$ for generic $n$. 

\section{Conclusions}

We investigated the relative entropy between the reduced density matrix, $\rho_A(t)$, and its long-time limit, $\rho_A(\infty)$, as a probe of the quantum Mpemba effect in quenches of closed quantum many-body systems. Various observables have been employed to examine the relaxation of a subsystem to equilibrium after a quench and to detect this phenomenon, including the entanglement asymmetry and the trace and Frobenius distances between $\rho_A(t)$ and $\rho_A(\infty)$. The entanglement asymmetry captures the Mpemba effect via the restoration of a global internal symmetry, and it cannot be applied when the final equilibrium state still breaks that symmetry. In contrast, the relative entropy and the distances between $\rho_A(t)$ and $\rho_A(\infty)$ do not suffer from this limitation; however, they are much more difficult to access in practice.

In this work, we demonstrated that the relative entropy between $\rho_A(t)$ and $\rho_A(\infty)$ can be accurately approximated by the difference in their entanglement entropies, which drastically simplifies the calculation. This approximation is exact when the time evolution has no conserved quantities. We analytically justified it in systems governed by a Hamiltonian with local interactions, drawing on previous results concerning the structure of their eigenstates. We also numerically verified its validity in long-range interacting spin Hamiltonians using exact diagonalization. Remarkably, under this approximation, this relative entropy corresponds to the entanglement asymmetry associated with the symmetry generated by the time evolution operator itself. Therefore, it effectively monitors the restoration of time-translation symmetry in a subsystem. 

While we have focused on the relative entropy in order to make connection to the entanglement asymmetry, one can also apply the same reasoning and approximation \eqref{eq:decomposition} to other metrics. In particular,  one can relate   the Frobenius distance between $\rho_A(t)$ and $\rho_{{\rm d},A}$ to the difference of R\'enyi entropies, 
 \begin{eqnarray}
 S^{(2)}[\rho_{{\rm d},A}]-S^{(2)}[\rho_A(t)]\simeq\log\Bigg\{1+\frac{ ||\rho_A(t)-\rho_{{\rm d},A}||^2_2}{||\rho_{{\rm d},A}||^2_2}\Bigg\}
 \end{eqnarray} 
which, as mentioned above, is exact in the absence of any conserved quantities. 

Our findings open the door to the direct application of the extensive results on entanglement entropy in quantum quenches to the study of the quantum Mpemba effect. Here we illustrate this in two paradigmatic cases: integrable systems, where we applied the quasiparticle picture, and random unitary circuits, where we used the membrane picture. In particular, we found that the difference in entanglement entropies of $\rho_A(t)$ and $\rho_A(\infty)$ detects the quantum Mpemba effect in the same regimes previously studied using the entanglement asymmetry or the distance between the two states. 

We conclude by emphasizing that the entanglement asymmetry and the relative entropy examined in this work, in principle, probe distinct aspects of the distance from equilibrium: the former probes the restoration of a global internal symmetry, while the latter captures the recovery of time-translational invariance.
As a result, it is conceivable that the Mpemba effect could manifest in one of these measures while being absent in the other.
However, as previously noted, across all natural quench protocols investigated so far, no instance of such a discrepancy has been observed.

\textbf{Acknowledgments}. The authors wish to thank J. Goold for illuminating discussions as well as B. Bertini, K. Klobas and S. Murciano for collaboration on closely related topics. All authors acknowledge support from European Union-NextGenerationEU, in the framework of the PRIN 2022 Project HIGHEST no. 2022SJCKAH\_002.

\end{document}